\documentclass[aps,prd,twocolumn,superscriptaddress,nofootinbib]{revtex4-2}

\usepackage{amsfonts}
\usepackage{amssymb,amsmath}
\usepackage{mathtools}
\usepackage{mathrsfs}
\usepackage{txfonts}
\usepackage{graphicx}
\usepackage{dcolumn}
\usepackage{multirow}
\usepackage{natbib}
\usepackage{comment}
\usepackage{color}
\usepackage[dvipsnames]{xcolor}
\usepackage{enumitem,soul}
\usepackage{orcidlink}
\usepackage{hyperref}
\hypersetup{colorlinks=true}
\hypersetup{linktoc=all, colorlinks, linkcolor={blue}, citecolor={blue}, urlcolor={blue}}

\graphicspath{{Images/}}
\newlength{\DefaultTabColSep}
\definecolor{darkgreen}{RGB}{27,130,45}
\definecolor{darkblue}{rgb}{0,0,0.3}
\definecolor{darkred}{rgb}{0.7,0,0}

\begin{document}

\title{Sound-Horizon-Independent Test of Cosmic Distance Duality Relation Using Artificial Neural Networks and Gaussian Processes}

\author{Bo-Hao Jiang}
\author{Guo-Jian Wang\orcidlink{0000-0003-0272-5032}}
\author{Tao Yang\orcidlink{0000-0002-2161-0495}}
\email{Corresponding author: yangtao@whu.edu.cn}
\affiliation{School of Physics and Technology, Wuhan University, Wuhan 430072, China}

\date{\today}

\begin{abstract}
The cosmic distance duality relation (CDDR), \(D_L(z)(1+z)^{-2}/D_A(z) \equiv 1\), is a fundamental relation in modern cosmology linking luminosity distance \(D_L\) and angular diameter distance \(D_A\). 
We define \(\eta(z) \equiv D_L(z)(1+z)^{-2}/D_A(z)\) and reconstruct the relevant quantities over \(0 < z < 2.5\) using two independent non-parametric methods: Gaussian processes (GP) and artificial neural networks (ANN). 
Specifically, we reconstruct \(D_L\) from three different Type Ia supernovae (SNe~Ia) compilations (PantheonPlus, Union3, and DES-Dovekie), the baryon acoustic oscillation distance ratio \(D_M/D_H\) from SDSS and DESI, and the Hubble function \(H(z)\) from cosmic chronometer (CC) data. 
Our CDDR test is independent of both the cosmological model and the sound horizon \(r_d\) calibration, as \(D_M/D_H\) naturally eliminates \(r_d\). 
We render our GP reconstructions insensitive to mean-function and kernel choices via full Bayesian marginalization over hyperparameters, with ANN as a cross-check---revealing the impact of data sparsity on the latter.
Our results show no significant deviation from the CDDR exceeding \(2\sigma\), and for the CC+BAO+DES-Dovekie combination it remains within \(1\sigma\) over most of the redshift range. 
With \(r_d\) eliminated, varying the only remaining external parameter \(M_B\) systematically shifts \(\eta(z)\) through the \(M_B\)--\(H_0\) degeneracy, indicating that the observed \(\eta\neq1\) signal is closely tied to \(H_0\) tension among datasets and is likely driven by \(M_B\) systematics rather than a true CDDR violation.
Future high-quality CC, BAO, and SNe~Ia data---with either improved \(M_B\) calibration or \(M_B\)-free distance measurements---will be essential for extending this calibration-free approach.
\end{abstract}

\maketitle

\section{Introduction}

The cosmic distance duality relation (CDDR) establishes a fundamental connection in modern cosmology between luminosity distance \(D_L(z)\) and angular diameter distance \(D_A(z)\) via \(D_L(z) = D_A(z) (1+z)^2\), assuming metric gravity, null geodesics, and photon conservation \cite{Bassett2004, Ellis2007}.
Any observed deviation from this relation would signal new physics or unaccounted systematics in the data \cite{Bassett2004, Bassett_2004, Uzan:2004my, Ellis_2007, Avgoustidis_2010, Eills_2013, Schuller:2017dfj, Azevedo:2021npm, Santos_2025}.

Recent years have seen extensive tests of the CDDR using a wide range of datasets and techniques \cite{Gao:2025ozb, Avila:2025sjz, Luo:2025vos, Yang:2025qdg, Li:2025htp, Xie:2025yun, Zheng:2025cgq, Qi:2024acx, Kanodia:2025jqh, Hu:2026yda, Tiwari:2026pzk}. 
Among these, the combination of Type Ia supernovae (SNe~Ia) \cite{Brout:2022vxf, DES:2024jxu, Rubin:2023jdq, DES:2025sig} and Baryon Acoustic Oscillations (BAO) \cite{Gil-Marin:2016wya, eBOSS:2020gbb, eBOSS:2020lta, DESI:2025zgx} has emerged as the most powerful and widely used probe, offering high precision, broad redshift coverage, and abundant data. 
Crucially, because the CDDR is independent of the cosmological model and expansion history, it provides a unique consistency check for BAO and SNe~Ia distance measurements \cite{Lopez-Hernandez:2025lbj, Woo:2026ice, Dinda:2025hiu, Dinda:2026uff, Afroz:2025iwo}. 
With the latest data, this test has been further extended to explore possible connections between CDDR violations and dynamical dark energy or the Hubble tension \cite{Tiwari:2026pzk, Teixeira:2025czm, Alfano:2025gie, Afroz:2025iwo}.

Testing the CDDR with BAO and SNe~Ia data typically requires calibration of the sound horizon~$r_d$ and the absolute magnitude~$M_B$, or priors on them \cite{Azevedo:2021npm, Kanodia:2025jqh, Alfano:2025gie, Teixeira:2025czm, Afroz:2025iwo, Tiwari:2026pzk}. 
To overcome the~$r_d$ dependence, we adopt the method of~Liu et al.~\cite{Liu:2024yib}, which uses cosmic chronometers (CC) to reconstruct~$H(z)$ and combines it with transverse and line-of-sight BAO data to derive~$D_A(z)$ independently of~$r_d$ and~$H_0$. 
For~$M_B$, we fix~$M_B$ to multiple values to assess its impact on~$\eta(z)$~\cite{Kanodia:2025jqh}.

A further challenge is the redshift mismatch between~$D_L$ and~$D_A$ measurements. Early attempts to match SNe~Ia to galaxy cluster redshifts used a simple interval $\Delta z < 0.005$~\cite{Holanda:2010vb, Li:2011exa}, later extended to binning~\cite{Meng_2012}, and more recently to a distance-based criterion $[D_C(z)-D_C(z_{\rm SNe~Ia})]/D_C(z) \leq 5\%$~\cite{Hu:2026yda}. 
To move beyond the limitations of such matching and enable comparisons at arbitrary redshifts, non-parametric reconstruction methods—particularly Gaussian processes (GP)~\cite{Keeley:2020aym, Gao:2025ozb, Avila:2025sjz, Luo:2025vos} and artificial neural networks (ANN)~\cite{Yang:2025qdg, Li:2025htp, Zheng:2025cgq, Qi:2024acx, Xie:2025yun}—have become widely adopted. 
Yet these methods are not without challenges. 
For GP, the choice of kernel and mean-function can introduce systematic biases comparable to those from different cosmological models~\cite{Ruchika:2025mkx, Hwang:2022hla}. 
ANN offers a fully data-driven alternative, reconstructing distance-redshift relations without assuming a functional form~\cite{Tang:2022ykd, Yang:2025qdg, Li:2025htp, Xie:2025yun}; however, its reconstruction may be less reliable in regions where data are sparse. 
Given their complementary strengths and distinct systematics, we employ both GP and ANN to provide a cross-checked reconstruction framework for testing the CDDR. 
We refer the interested reader to Refs.~\cite{Cui:2025rri, Luo:2025vos, Ruchika:2025mkx, Wang:2019vxv} for a more comprehensive discussion.

Building on these considerations, in this work we adopt a calibration-independent approach that offers several methodological innovations. 
Following Ref.~\cite{Liu:2024yib}, we combine CC with BAO measurements to derive~$D_A(z)$ without any prior on~$r_d$ or~$H_0$, by directly reconstructing~$H(z)$ and the observed BAO ratio~$R_{MH}(z)$. 
We employ both an improved GP method---with full Bayesian marginalization over hyperparameters~\cite{Ruchika:2025mkx, Hwang:2022hla}---and ANN~\cite{Wang:2019vxv,Gadbail:2026gkx} to reconstruct $H(z)$, $R_{MH}(z)$, and $D_L(z)$. 
From these reconstructions we obtain \(\eta(z)\) over~$0<z<2.5$. We find no deviation exceeding \(2\sigma\) from the CDDR, and show that the \(\eta\neq1\) signal likely points to \(H_0\) tension and \(M_B\) calibration systematics rather than new physics.

\section{Data and Methodology}
\label{sec:data_method}

%\subsection{Reconstruction Framework}
%\label{subsec:framework}

The core of our analysis is a model-independent evaluation of the CDDR, encapsulated in the following equation:
\begin{equation}
\label{eq:eta_core}
\eta(z) \equiv \frac{D_L(z)}{D_A(z) (1+z)^2} = \frac{H^{\text{CC}}(z) \cdot D_L^{\text{SNe~Ia}}(z)}{(1+z) c \cdot R^{\text{BAO}}_{MH}(z)}.
\end{equation}
This expression follows directly from the definitions \(D_H = c/H\), \(D_A = D_M/(1+z)\), and \(R_{MH} \equiv D_M/D_H\). Crucially, the dimensionless ratio~$R_{MH}\equiv (D_M/r_d)/(D_H/r_d) = D_M/D_H$ is independent of~$r_d$, eliminating the need for a CMB prior. 

To compute $\eta(z)$ via Eq.~\eqref{eq:eta_core}, we reconstruct the three ingredients---$H(z)$ from CC, $R_{MH}(z)$ from BAO, and $D_L(z)$ from SNe~Ia---using GP and ANN.
At $z=0$, we include $(0,0)$ as a virtual data point with effectively zero uncertainty (numerically implemented in the code as $\sigma=10^{-6}$) for both $D_L$ and $R_{MH}$, justified by the FLRW metric ($D_M(0)=0$). This constraint prevents unphysical divergence of $\eta(z)$ as $z\to0$ and improves low-redshift stability. It is imposed only in the reconstruction process and does not represent an actual observation, consistent with our aim to test the CDDR within the FLRW framework, not the FLRW model itself \cite{Dinda:2025hiu}.
For $D_L(z)$, we reconstruct it directly rather than the distance modulus~$\mu(z)$ to avoid unphysical negative derivatives~\cite{Hwang:2022hla, Wu:2026klh, Dinda:2025hiu, Dinda:2026uff}. 

\subsection{Observational data}

\paragraph{Type Ia supernovae}
We use three compilations: PantheonPlus (PP)~\cite{Brout:2022vxf}~\footnote{\url{https://github.com/PantheonPlusSH0ES/DataRelease}}, Union3 (U3)~\cite{Rubin:2023jdq}~\footnote{\url{https://github.com/rubind/union3_release}}, and DES-Dovekie (DD)~\cite{DES:2025sig}~\footnote{\url{https://github.com/des-science/DES-SN5YR}}. 
The PP dataset consists of 1701 light curves from 1550 spectroscopically confirmed SNe~Ia over $0.001 < z < 2.27$. 
The U3 compilation includes 2087 events over the same redshift range, at 22 nodes in the range $0.05 \leq z < 2.27$. 
The DD catalogue follows from a re-analysis of the Dark Energy Survey (DES) 5-year sample of SNe~Ia (DES-SN5YR)~\cite{DES:2024jxu}, including a total of 1820 SNe~Ia in the range $0.02 < z < 1.15$. 
Because the DES supernovae are photometrically classified, the distance uncertainties are renormalized by the Bayesian Estimation Applied to Multiple Species (BEAMS) probability of being Type Ia~\cite{DES:2025sig, DES:2024jxu, DES:2024hip}, which greatly enlarges the uncertainties of likely contaminants and thereby reduces their influence on the cosmological analysis~\cite{Wu:2026klh}. As a result, a small fraction of the DD data points carry substantially larger uncertainties than the rest of the sample.
For each dataset, we extract the distance modulus~$\mu$ and convert it to~$D_L(z)$ via $D_L = 10^{\mu/5 - 5} \text{Mpc}$. 

\paragraph{Baryon Acoustic Oscillations}
To obtain $R_{MH}$, we utilize BAO measurements of $D_M/r_d$ and $D_H/r_d$ from Dark Energy Spectroscopic Instrument (DESI) Data Release 2 (DR2)~\cite{DESI:2025zgx} and the Sloan Digital Sky Survey (SDSS), which includes DR12~\cite{Gil-Marin:2016wya} and DR16~\cite{eBOSS:2020lta, eBOSS:2020gbb}~\footnote{\url{https://github.com/CobayaSampler/bao_data}.}. 
We reconstruct $R_{MH}(z)$ separately for DESI and SDSS, rather than merging them into a single dataset. 
This is because, despite coming from different surveys, the DESI DR2 and SDSS footprints have significant overlap, leading to a non-negligible correlation between the two datasets. 
Estimates of this correlation give \(C \sim 0.57\) for overlapping redshift bins \cite{DESI:2025zgx}. 
Merging them would underestimate the combined uncertainty. 
By keeping the reconstructions separate, we preserve the independence of the two datasets and use the differences between their results as a cross-check for systematic uncertainties, with any discrepancy being well below the \(2\sigma\) level \cite{DESI:2025zgx}. 
For Gaussian measurements, we propagate errors from the joint covariance matrix. 
For the non-Gaussian eBOSS ELG at \(z=0.845\), we derive \(R_{MH}\) via Monte Carlo sampling from the published grid likelihood, taking the median and 16th/84th percentiles as the estimate and \(1\sigma\) interval, and symmetrize the resulting asymmetric uncertainties by averaging the upper and lower \(1\sigma\) values.
This simplification is applied to all \(R_{MH}\) measurements, so that the uncertainties reported in Table~\ref{tab:bao_ratio} are Gaussian and symmetric, which is convenient for GP and ANN.
The results are listed in Table~\ref{tab:bao_ratio}.
\begin{table}[htb]
\centering
\begin{tabular}{ccc}
\hline
\textbf{Survey/Tracer} & $z_{\mathrm{eff}}$ & $R_{\mathrm{MH}}$ \\
\hline
\multicolumn{3}{c}{\text{DESI}} \\
\hline
LRG1 & 0.51 & $0.621 \pm 0.017$ \\
LRG2 & 0.706 & $0.892 \pm 0.021$ \\
LRG3+ELG1 & 0.934 & $1.223 \pm 0.019$ \\
ELG2 & 1.321 & $1.947 \pm 0.045$ \\
QSO & 1.484 & $2.381 \pm 0.135$ \\
Ly-$\alpha$ & 2.33 & $4.517 \pm 0.096$ \\
\hline
\multicolumn{3}{c}{\text{SDSS}} \\
\hline
LRG (DR12) & 0.38 & $0.413 \pm 0.012$ \\
LRG (DR12) & 0.51 & $0.597 \pm 0.017$ \\
LRG (DR16) & 0.698 & $0.893 \pm 0.029$ \\
ELG (Grid) & 0.845 & $1.009 \pm 0.136$ \\
QSO & 1.48 & $2.284 \pm 0.098$ \\
Ly-$\alpha$ auto & 2.334 & $4.207 \pm 0.267$ \\
Ly-$\alpha$ cross & 2.334 & $4.130 \pm 0.263$ \\
\hline
\end{tabular}
\caption{BAO distance ratio $R_{\mathrm{MH}} = D_M/D_H$.}
\label{tab:bao_ratio}
\end{table}

\paragraph{Cosmic Chronometers}
We compile 34 CC $H(z)$ measurements spanning $0.07 \leq z \leq 1.965$ from independent studies~\cite{Jimenez:2003iv, Simon:2004tf, Stern:2009ep, Moresco:2012jh, Zhang:2012mp, Moresco:2015cya, Moresco:2016mzx, Ratsimbazafy:2017vga, Moresco:2022phi, Jiao:2022aep, Tomasetti:2023kek, Loubser:2025snu, Niu:2025tzp}. 
The reported uncertainties are total errors combining statistical and systematic contributions~\cite{Moresco:2020fbm}. 
In addition, as noted in the appendix of~\cite{Niu:2025tzp}, we apply the correction to the $H(0.09)$ value from~\cite{Jimenez:2003iv}, updating it from $69 \pm 12$ to $70.7 \pm 12.3$ km/s/Mpc. 
All uncertainties are treated as Gaussian in our analysis and we incorporate the full covariance matrix~\footnote{\url{https://gitlab.com/mmoresco/CCcovariance}} in the reconstruction.. The full list of CC data used in this work is presented in Table~\ref{tab:cc_data}.
\begin{table}[htbp]
\centering
\begin{tabular}{ccc}
\hline
$z$ & $H(z)$ [km/s/Mpc] & References \\
\hline
0.07 & $69.0 \pm 19.6$ & Zhang et al.~\cite{Zhang:2012mp} \\
0.09 & $70.7 \pm 12.3$ & Jimenez et al.~\cite{Jimenez:2003iv}, Niu et al.~\cite{Niu:2025tzp} \\
0.12 & $68.6 \pm 26.2$ & Zhang et al.~\cite{Zhang:2012mp} \\
0.17 & $83.0 \pm 8.0$ & Simon et al.~\cite{Simon:2004tf} \\
0.1791 & $75.0 \pm 4.0$ & Moresco et al.~\cite{Moresco:2012jh} \\
0.1993 & $75.0 \pm 5.0$ & Moresco et al.~\cite{Moresco:2012jh} \\
0.20 & $72.9 \pm 29.6$ & Zhang et al.~\cite{Zhang:2012mp} \\
0.27 & $77.0 \pm 14.0$ & Simon et al.~\cite{Simon:2004tf} \\
0.28 & $88.8 \pm 36.6$ & Zhang et al.~\cite{Zhang:2012mp} \\
0.3519 & $83.0 \pm 14.0$ & Moresco et al.~\cite{Moresco:2012jh} \\
0.3802 & $83.0 \pm 13.5$ & Moresco et al.~\cite{Moresco:2016mzx} \\
0.40 & $95.0 \pm 17.0$ & Simon et al.~\cite{Simon:2004tf} \\
0.4004 & $77.0 \pm 10.2$ & Moresco et al.~\cite{Moresco:2016mzx} \\
0.4247 & $87.1 \pm 11.2$ & Moresco et al.~\cite{Moresco:2016mzx} \\
0.44497 & $92.8 \pm 12.9$ & Moresco et al.~\cite{Moresco:2016mzx} \\
0.47 & $89.0 \pm 49.6$ & Ratsimbazafy et al.~\cite{Ratsimbazafy:2017vga} \\
0.4783 & $80.9 \pm 9.0$ & Moresco et al.~\cite{Moresco:2016mzx} \\
0.48 & $97.0 \pm 62.0$ & Stern et al.~\cite{Stern:2009ep} \\
0.50 & $72.1 \pm 34.7$ & Loubser et al.~\cite{Loubser:2025snu} \\
0.5929 & $104.0 \pm 13.0$ & Moresco et al.~\cite{Moresco:2012jh} \\
0.6797 & $92.0 \pm 8.0$ & Moresco et al.~\cite{Moresco:2012jh} \\
0.7812 & $105.0 \pm 12.0$ & Moresco et al.~\cite{Moresco:2012jh} \\
0.80 & $113.1 \pm 25.22$ & Jiao et al.~\cite{Jiao:2022aep} \\
0.8754 & $125.0 \pm 17.0$ & Moresco et al.~\cite{Moresco:2012jh} \\
0.88 & $90.0 \pm 40.0$ & Stern et al.~\cite{Stern:2009ep} \\
0.90 & $117.0 \pm 23.0$ & Simon et al.~\cite{Simon:2004tf} \\
1.037 & $154.0 \pm 20.0$ & Moresco et al.~\cite{Moresco:2012jh} \\
1.26 & $135.0 \pm 65.0$ & Tomasetti et al.~\cite{Tomasetti:2023kek} \\
1.30 & $168.0 \pm 17.0$ & Simon et al.~\cite{Simon:2004tf} \\
1.363 & $160.0 \pm 33.6$ & Moresco~\cite{Moresco:2015cya} \\
1.43 & $177.0 \pm 18.0$ & Simon et al.~\cite{Simon:2004tf} \\
1.53 & $140.0 \pm 14.0$ & Simon et al.~\cite{Simon:2004tf} \\
1.75 & $202.0 \pm 40.0$ & Simon et al.~\cite{Simon:2004tf} \\
1.965 & $186.5 \pm 50.4$ & Moresco~\cite{Moresco:2015cya} \\
\hline
\end{tabular}
\caption{CC $H(z)$ measurements used in this analysis.}
\label{tab:cc_data}
\end{table}

\subsection{Generalized Gaussian Process}
\label{subsec:GP}

A GP is fully specified by its mean function $\mu(z)$ and covariance kernel $k(z,\tilde{z})$ \cite{Seikel:2012uu}:
\begin{equation}
f(z) \sim \mathcal{GP}\left( \mu(z), k(z,\tilde{z}) \right).
\end{equation}
We consider the Mat\'{e}rn family:
\begin{equation}
k_{\mathrm{M}}(z, \tilde{z}) = \sigma_f^2 \frac{2^{1-\nu}}{\Gamma(\nu)} \left( \frac{\sqrt{2\nu}|z-\tilde{z}|}{\ell} \right)^{\nu} K_{\nu}\left( \frac{\sqrt{2\nu}|z-\tilde{z}|}{\ell} \right),
\end{equation}
where \(\sigma_f\), \(\ell\), and \(\nu\) control the amplitude, correlation length, and smoothness, and \(K_{\nu}\) is the modified Bessel function of the second kind. 
The squared exponential kernel is recovered as \(\nu\to\infty\). 

We follow the Gen GP framework of Ref.~\cite{Ruchika:2025mkx}, which extends the full marginalization approach of Ref.~\cite{Hwang:2022hla} by treating the Mat\'ern smoothness parameter $\nu$ as an additional hyperparameter to be sampled via MCMC. 
This fully Bayesian treatment marginalizes over all hyperparameters, propagating their uncertainties into the final reconstruction and avoiding the potential biases associated with kernel selection. 

We adopt a $\Lambda$CDM mean function with hyperparameters $H_{0,\rm hyper}$ and $\Omega_{m,\rm hyper}$ under uniform priors $[50,90]$ km/s/Mpc and $[0.1,0.5]$, respectively. 
We omit the ``hyper'' subscript when there is no ambiguity; the reader should distinguish the mean-function hyperparameter $H_{0,\rm hyper}$ from the physical Hubble constant $H_0$. 
The kernel hyperparameters $\ell$, $\sigma_f$, and $\nu$ are assigned log-uniform priors: $\log_{10}\ell \sim \mathcal{U}[-5,5]$, $\log_{10}\sigma_f \sim \mathcal{U}[-5,5]$, and $\log_{10}\nu \sim \mathcal{U}[-2,1]$. 
These broad priors allow MCMC to fully explore the hyperparameter space. 

The hyperparameters are sampled from the posterior defined by the GP log-marginal likelihood:
\begin{equation}
\ln \mathcal{L} = -\frac{1}{2} \mathbf{y}^T K_y^{-1} \mathbf{y} - \frac{1}{2} \ln |K_y| - \frac{n}{2} \ln(2\pi),
\end{equation}
where $K_y = K(X,X) + C$ is the total covariance matrix including observational noise. 
Convergence is assessed by the Gelman-Rubin statistic. 
Achieving \(\hat{R} < 1.01\) would require far longer chains due to the enlarged parameter space. 
We adopt \(\hat{R} < 1.05\) as a practical threshold, and have verified that this choice changes the reconstructed \(\eta(z)\) by less than \(0.5\%\) of the \(1\sigma\) uncertainty. 

From the posterior distribution, we generate 1000 GP realizations, and take the median and the 16th/84th percentiles as the best-fit and $1\sigma$ confidence interval. 
The MCMC sampling uses \texttt{emcee}~\cite{foreman-mackey_emcee_2013}, and GP predictions use \texttt{GaPP}~\cite{Seikel:2012uu}\footnote{\url{https://github.com/lighink/GaPP3}}. For further details on GP methods, we refer the reader to Refs.~\cite{Seikel:2012uu,Hwang:2022hla,Ruchika:2025mkx}.

\subsection{Artificial Neural Networks}
\label{subsec:ANN}
\begin{figure*}
    \centering
    \includegraphics[width=1\linewidth]{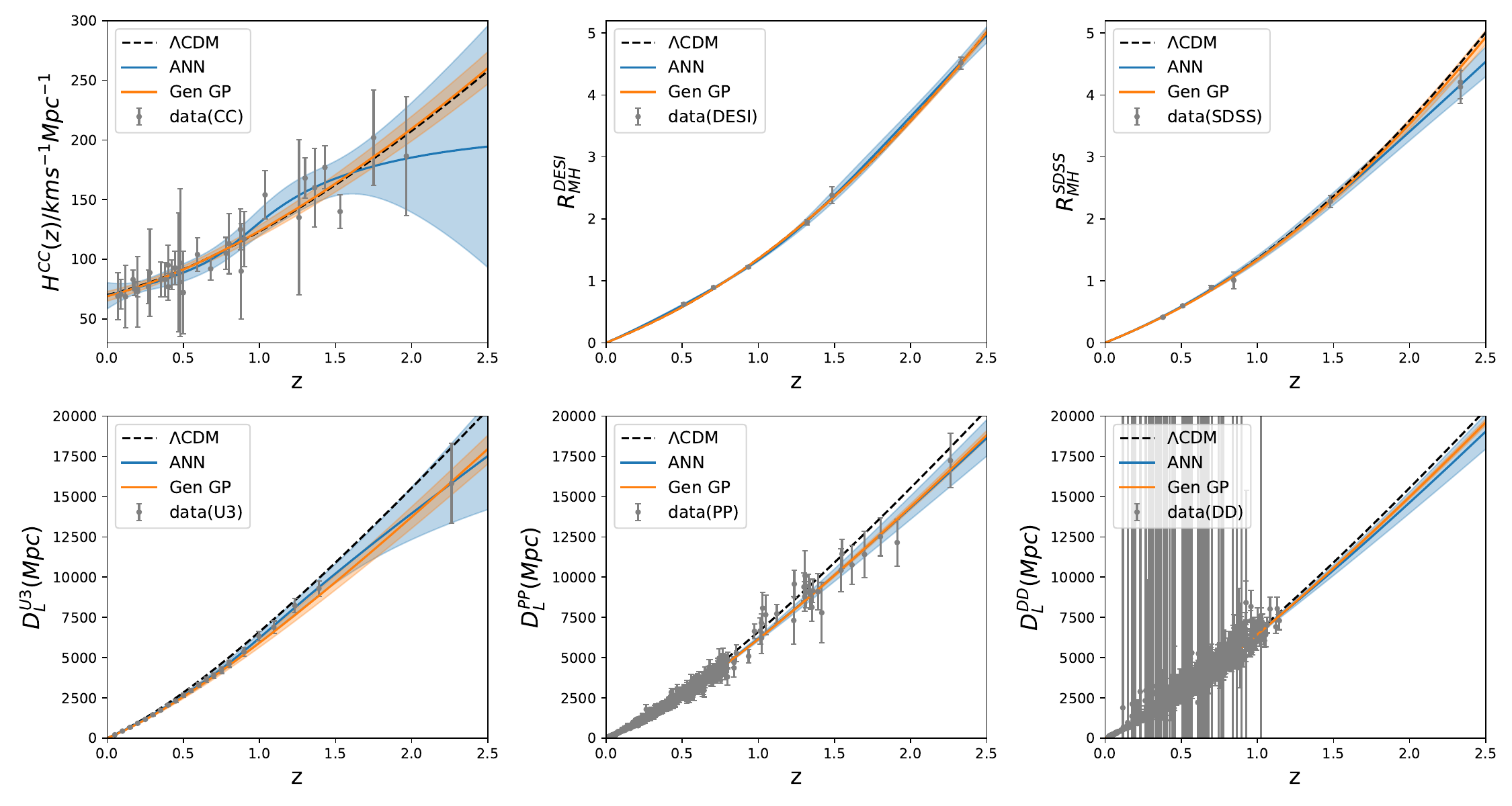}
    \caption{Reconstructed $H(z)$, $R_{MH}(z)$, and $D_L(z)$ from ANN and Gen GP. The black dashed line shows the fiducial $\Lambda$CDM model ($H_0=70$ km/s/Mpc, $\Omega_m=0.3$), and the grey points with error bars represent the observational data. Shaded regions indicate $1\sigma$ confidence intervals.}
    \label{fig:reconstruction_mcmcgp_multiann}
\end{figure*}
\begin{figure*}
    \centering
    \includegraphics[width=1\linewidth]{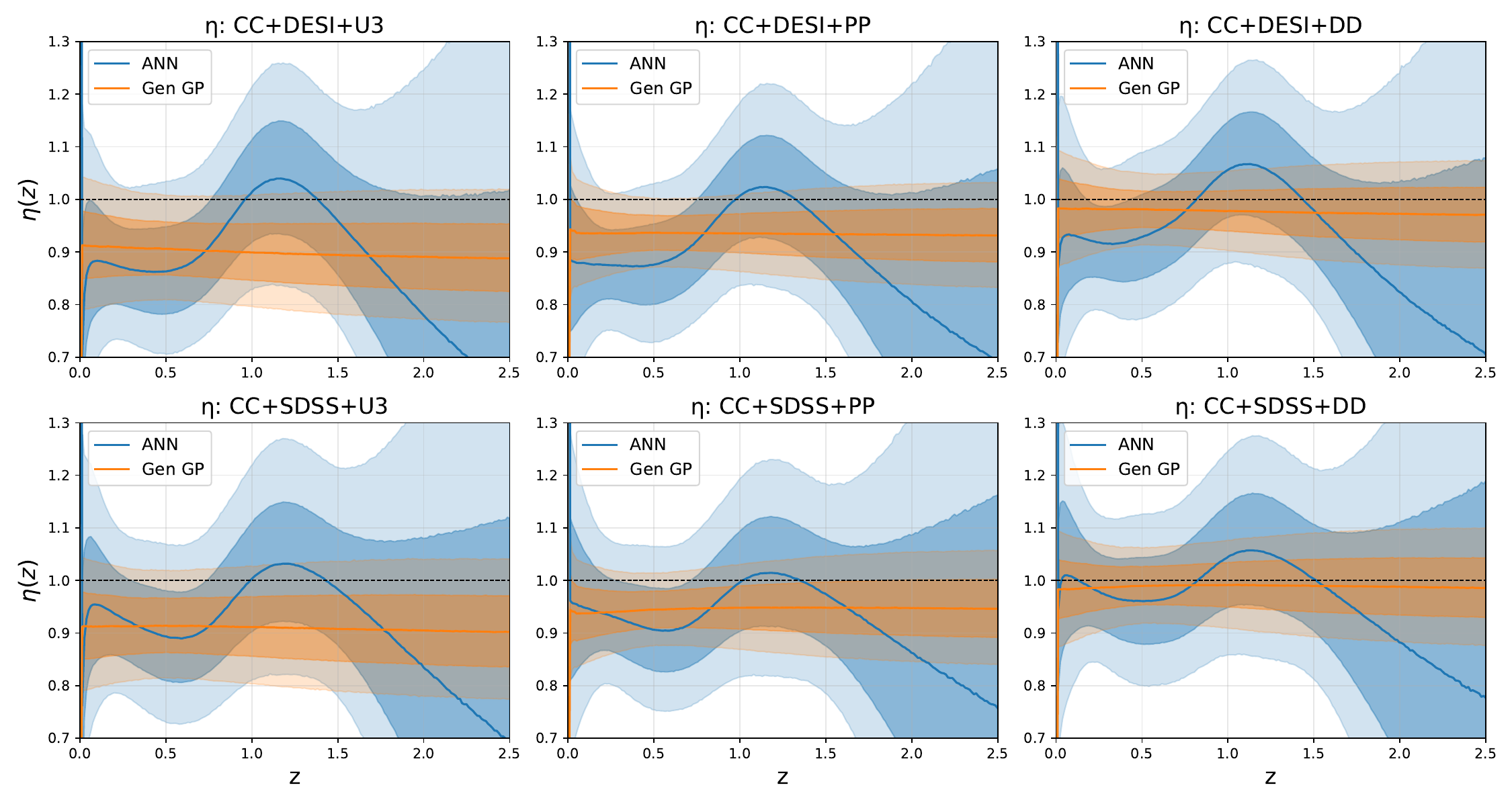}
    \caption{Reconstructed $\eta(z)$ from Figure~\ref{fig:reconstruction_mcmcgp_multiann} for six data combinations. The horizontal dashed line marks $\eta=1$. Dark and light shaded regions represent $1\sigma$ and $2\sigma$ confidence intervals, respectively.}
    \label{fig:eta_mcmcgp_multiann}
\end{figure*}
We use the publicly available \texttt{RefANN} package~\cite{Wang:2019vxv}\footnote{\url{https://github.com/Guo-Jian-Wang/refann}.} for the ANN reconstruction. 
ANN is fully data-driven, requiring no assumption of a specific functional form or Gaussian uncertainties, and scales linearly with dataset size rather than \(\mathcal{O}(N^3)\) as GP does, making it well-suited for large samples such as PP and DD \cite{Wang:2019vxv, Cui:2025rri}.
This method has been successfully applied to reconstruct the Hubble parameter, distance moduli, and cosmological parameters \cite{Wang:2019vxv, Yang:2025qdg, Cui:2025rri}.

Starting from the observed data and its covariance matrix, we generate 1000 simulated samples from the multivariate Gaussian distribution defined by the data covariance matrix. For each sample, we train an ANN with a different random initialization. 
Following Refs.~\cite{Wang:2019vxv, Gadbail:2026gkx}, we use a feedforward ANN with a single hidden layer of 1024 neurons and the Exponential Linear Unit (ELU, $\alpha=1$) activation function:
\begin{equation}
\text{ELU}(x) = \begin{cases}
x, & x \geq 0 \\
\alpha(e^x - 1), & x < 0
\end{cases}
\end{equation}
ELU mitigates the vanishing gradient problem and accelerates convergence compared to ReLU or sigmoid \cite{Cui:2025rri}. 
Both inputs and targets are standardized using z-score normalization. 
Training minimizes the L1 loss (mean absolute error):
\begin{equation}
\mathcal{L}_1 = \frac{1}{N} \sum_{i=1}^{N} |\hat{y}_i - y_i|,
\end{equation}
which is more robust to outliers than mean squared error \cite{Wang:2019vxv}. 
We use the Adam optimizer with an initial learning rate of $10^{-2}$ that decays during training, and train for 30,000 iterations to ensure convergence. 
A weight decay of $5\times10^{-4}$ is applied for regularization. 
Batch normalization is used for the CC $H(z)$ reconstruction, but omitted for $D_L(z)$ and $R_{MH}(z)$ where it provides no improvement. 

After training the ensemble of 1000 networks, we evaluate each at a common set of redshift points, yielding 1000 reconstructed function values at each redshift~\cite{Gadbail:2026gkx}. 
The mean and standard deviation of the ensemble are taken as the best-fit reconstruction and $1\sigma$ uncertainty. The full covariance matrix is also obtained, providing correlations between predictions at different redshifts. 
This ensemble approach propagates observational uncertainties through the reconstruction without relying on analytic error propagation.

\section{Results and Discussion}
\label{sec:results}

We reconstruct $H(z)$, $R_{MH}(z)$, and $D_L(z)$ using Gen GP and ANN, and combine them via Eq.~\eqref{eq:eta_core} to obtain $\eta(z)$. As described in Sec.~\ref{sec:data_method}, the origin constraints $(0,0)$ are imposed for both $D_L$ and $R_{MH}$.
Figures~\ref{fig:reconstruction_mcmcgp_multiann} and~\ref{fig:eta_mcmcgp_multiann} present the reconstructions and the corresponding $\eta(z)$ for six data combinations: CC+DESI+U3, CC+DESI+PP, CC+DESI+DD, CC+SDSS+U3, CC+SDSS+PP, and CC+SDSS+DD.

\subsection{Overall consistency with the CDDR}
\label{subsec:overall}

Figure~\ref{fig:reconstruction_mcmcgp_multiann} shows the reconstructed $H(z)$, $R_{MH}(z)$, and $D_L(z)$. 
Both Gen GP and ANN reconstructions agree well with the observational data, with Gen GP yielding smoother curves and ANN showing localized fluctuations driven by the data. The ANN uncertainties are larger than those of Gen GP across all three quantities. 

Figure~\ref{fig:eta_mcmcgp_multiann} shows the resulting $\eta(z)$ for six data combinations. 
Across all combinations, no deviation from $\eta=1$ exceeds $2\sigma$. The Gen GP reconstructions are consistently smooth and nearly constant, with central values between approximately $0.9$ and $1$. 
For the CC+BAO+DD combinations, $\eta=1$ lies within the $1\sigma$ confidence interval over $0<z<2.5$, indicating the strongest consistency with the CDDR. 

The ANN reconstructions of $\eta(z)$ reveal two behaviors. First, they exhibit a broad peak around $z\sim1.2$ and a subsequent downturn at $z>1.5$, both remaining within the $1\sigma$ deviation from the CDDR in all cases. The peak and downturn coincide with the behavior of the ANN-reconstructed $H(z)$ (Fig.~\ref{fig:reconstruction_mcmcgp_multiann}, upper left panel), where a slight peak and a mild downturn at higher redshifts are present. This correspondence indicates that these structures in $\eta(z)$ are driven by the ANN reconstruction of the CC $H(z)$.

Second, at low redshifts there is a slight difference between the DESI-based and SDSS-based $\eta(z)$ reconstructions. 
For the CC+SDSS+SNe~Ia (lower row of Fig.~\ref{fig:eta_mcmcgp_multiann}), the $1\sigma$ deviation from $\eta=1$ is largely confined to $z\sim0.5$, while the DESI-based $\eta(z)$ (upper row) remains below unity at the $1\sigma$ level down to $z\sim0.2$. 
This difference reflects the distinct low-redshift BAO constraints from the two surveys: SDSS provides two LRG measurements at $z=0.38$ and $z=0.51$, whereas DESI's lowest BAO point is at $z=0.51$, leaving the $z<0.5$ region less constrained.

\subsection{Impact of data sparsity on ANN reconstructions}
\label{subsec:diagnosis}
\begin{figure*}
    \centering
    \includegraphics[width=1\linewidth]{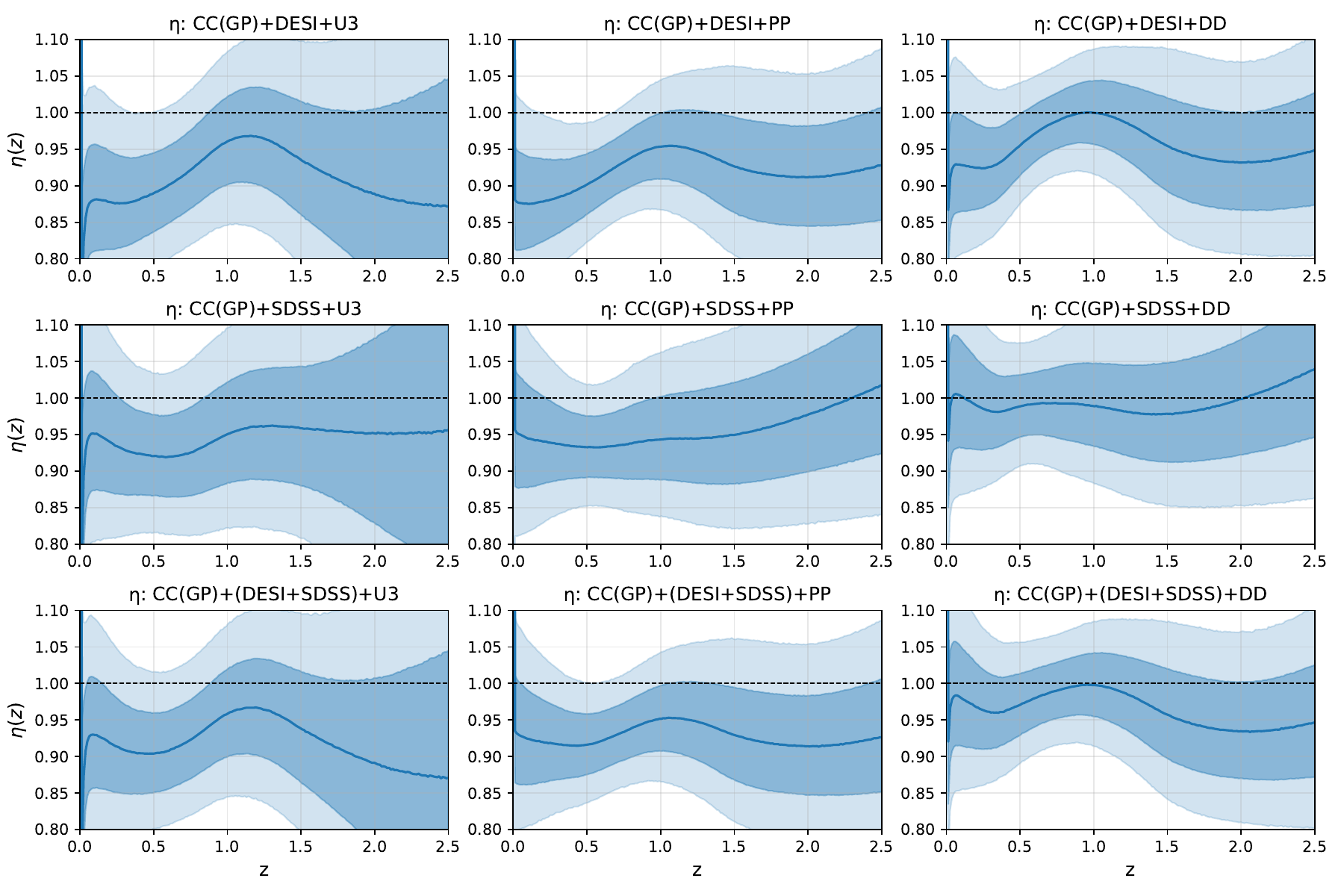}
    \caption{Reconstructed \(\eta(z)\) obtained by combining Gen GP reconstructed \(H(z)\) with ANN-reconstructed \(R_{MH}(z)\) and \(D_L(z)\), for three BAO datasets (DESI, SDSS, and DESI+SDSS) and three SNe~Ia datasets.}
    \label{fig:eta_multiann_gp_CC}
\end{figure*}
\begin{figure*}
    \centering
    \includegraphics[width=1\linewidth]{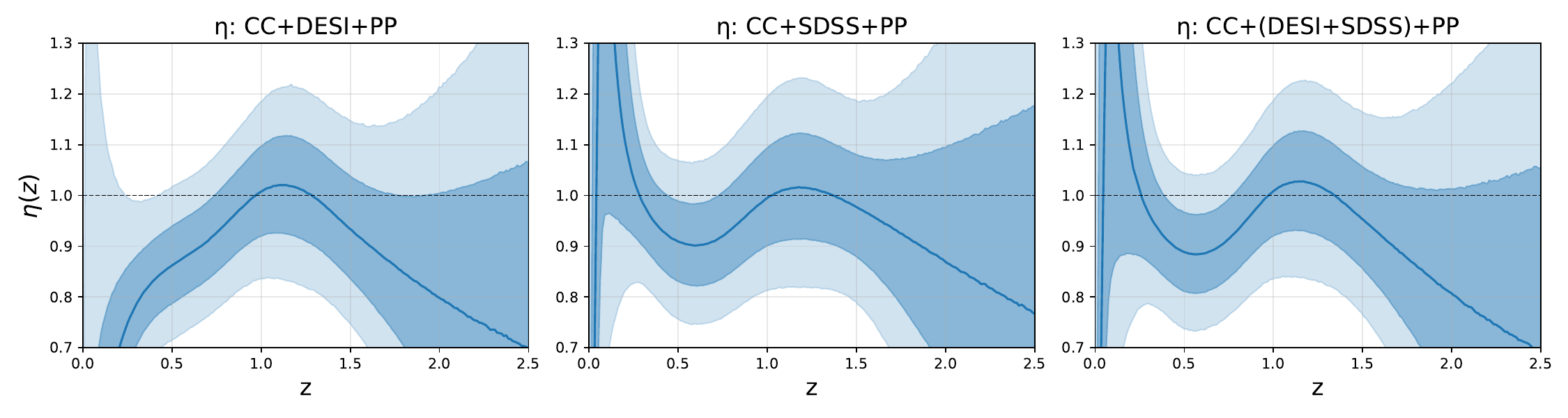}
    \caption{Reconstructed \(\eta(z)\) from ANN without imposing the \((0,0)\) anchor, i.e., \(R_{MH}(0)=0\), using the PP compilation.}
    \label{fig:eta_multiann_nonzero}
\end{figure*}

ANN is a purely data-driven method: the features it produces may either reflect genuine structures in the underlying data or arise from overfitting to sparse and noisy measurements. 
The two ANN behaviors described above---the broad peak and downturn, and the low-redshift difference between BAO samples---do not indicate a violation of the CDDR, as they are consistent with $\eta=1$ within the uncertainties. 
Rather, they reflect known limitations of the current data: the sparsity and noise of CC measurements at high redshifts, and the lack of BAO data below $z=0.51$ for DESI. 
To illustrate how these limitations affect the reconstruction, we perform two tests.

First, we replace the ANN-reconstructed CC $H(z)$ with its Gen GP counterpart, while keeping the ANN-reconstructed $R_{MH}(z)$ and $D_L(z)$ unchanged. 
Second, we construct a combined BAO sample, denoted DESI+SDSS, which consists of all six DESI BAO points plus the SDSS LRG measurement at $z=0.38$, to assess how an additional low-redshift BAO data point affects the reconstruction. 
Figure~\ref{fig:eta_multiann_gp_CC} shows the resulting $\eta(z)$.

Replacing the ANN-reconstructed CC $H(z)$ with the Gen GP version largely removes the broad peak and downturn. 
The central values around $z\sim1.2$ are now mostly below unity, and the downturn at $z>1.5$ is suppressed, giving way to a flat or even slightly rising trend at high redshifts for the CC+SDSS+PP and CC+SDSS+DD combinations. 
This shows that the peak and downturn in the pure-ANN $\eta(z)$ are sensitive to the CC reconstruction method, and are likely amplified by the sparsity and scatter of the CC data at $z>1$, where only a few measurements with large uncertainties exist and none beyond $z=1.965$. 
Future CC measurements at high redshifts will therefore be essential for reducing this sensitivity. 

The low-redshift behavior, by contrast, is governed by the BAO coverage. As shown in Fig.~\ref{fig:eta_multiann_gp_CC}, the CC+SDSS+U3 and CC+SDSS+PP combinations still show a $1\sigma$ excess above unity near $z\sim0.5$. 
The DESI-based combinations exhibit a broader deviation from $\eta=1$ than their SDSS counterparts; notably, DESI+PP exceeds the $2\sigma$ threshold near $z\sim0.5$. 
When the SDSS point at $z=0.38$ is added to the DESI sample (DESI+SDSS), the low-redshift $\eta(z)$ shifts closer to the SDSS-based results, and the CC+DESI+SDSS+PP combination reduces the $z\sim0.5$ excess to the boundary of the $2\sigma$ interval. 
This improvement demonstrates that even a single additional low-redshift BAO data point can noticeably tighten the constraints. 

In addition, we examine the effect of the anchor. When no anchor is imposed to constrain the BAO reconstruction, the result (CC+BAO+PP) is shown in Fig.~\ref{fig:eta_multiann_nonzero}. 
We show the PP case because it contains a large number of low-redshift data extending down to \(z\sim0.001\); consequently, the PP reconstruction of \(D_L(z)\) is the least sensitive to the anchor. 
Therefore, this case allows us to isolate the effect of the anchor on the BAO reconstruction.
Although the CDDR is still satisfied within roughly \(2\sigma\), as \(z\to0\) the central values of \(\eta(z)\) either drop directly to zero or diverge and then approach zero from negative values, both of which are unphysical. 
This exercise highlights that low-redshift BAO data are essential for stabilizing the reconstruction at low \(z\), and future DESI data releases covering \(z<0.5\) will therefore be valuable for improving the low-redshift reconstruction.

\subsection{Connection to the Hubble tension and absolute magnitude calibration}
\label{subsec:H0_MB}

Beyond the data-driven effects discussed in Sec.~\ref{subsec:diagnosis}, Figs.~\ref{fig:eta_mcmcgp_multiann} and~\ref{fig:eta_multiann_gp_CC} also show that the mean $\eta(z)$ depends systematically on the choice of SNe~Ia compilation: the CC+BAO+DD combination deviates least from the CDDR, while CC+BAO+U3 deviates most. 
This feature is independent of the BAO sample and the reconstruction method, pointing to an intrinsic difference among the SNe~Ia datasets. We now examine the physical origin of this trend. 

Tiwari et al.~\cite{Tiwari:2026pzk} recently showed, using a joint analysis of uncalibrated BAO (DESI DR2) and SNe~Ia (PantheonPlus) data under the CDDR assumption, that the combination of the SH0ES absolute magnitude $M_B$ calibration and the Planck $\Lambda$CDM-inferred $r_d$ lies $7\sigma$ away from the CDDR consistency region in the $r_d$--$M_B$ plane. 
In their framework, the BAO+SNe~Ia data constrain only the combinations $$\mathcal{A}=M_B + 5\log_{10}\left(\eta \frac{r_d}{\text{Mpc}}\right)=m(z)+5\log_{10}\left(\frac{\theta_{BAO}}{1+z}\right)-25$$ and $\mathcal{M}=M_B - 5\log_{10}(H_0/\text{(km/s/Mpc)})$, where $m(z)$ is the observed apparent magnitude and $\theta_{BAO}(z)$ is the observed angular scale of the BAO feature. 
For fixed $r_d$, $\mathcal{A}$ and $\mathcal{M}$, any shift in $H_0$ must be compensated by a violation of the CDDR, with $\delta H_0/H_0 = -\delta\eta/\eta$. 
Reconciling the $H_0$ tension between $67$ and $73$ km/s/Mpc would therefore require an $\mathcal{O}(8\text{--}10\%)$ deviation from $\eta=1$.

To test whether the $\eta\neq1$ signal we observe is connected to this mechanism, we compare the $H_0$ values preferred by each SNe~Ia compilation with the CC-only constraint. 
Table~\ref{tab:eta_means} provides a quantitative test using our GP-reconstructed $\eta(z)$. 
The relative $H_0$ deviations ($\delta H_0/H_0^{\rm CC}$, with $H_0^{\rm CC} \sim 69.1$ km/s/Mpc from a direct $\Lambda$CDM fit to the CC data) are closely tracked by the corresponding $\eta$ deviations, $\delta\eta \equiv 1 - \langle\eta\rangle$, where $\langle\eta\rangle$ is the mean of $\eta(z)$ over $0<z<2.5$. 
U3, which favors the highest $H_0$, produces the largest $\eta$ deviation ($\sim 10\%$); DD, whose $H_0$ is closest to the CC value, yields $\eta$ closest to unity ($\sim 2\%$). 
The two sets of deviations are in good agreement, supporting a common origin in the $H_0$ tension.

\begin{table}[htbp]
\centering
\begin{tabular}{cccc}
\hline
\text{SNe~Ia} & $\delta H_0 / H_0^{\rm CC}$ & $\delta\eta$ (SDSS) & $\delta\eta$ (DESI) \\
 & (\%) & (\%) & (\%) \\
\hline
U3 & +9.6 & $9.5$ & $10.2$ \\
PP & +6.4 & $5.9$ & $7.2$ \\
DD & +1.6 & $1.5$ & $2.8$ \\
\hline
\end{tabular}
\caption{Comparison of $H_0$ relative deviations (relative to CC-only) and $\eta$ deviations from unity.}
\label{tab:eta_means}
\end{table}
\begin{figure}
    \centering
    \includegraphics[width=0.75\linewidth]{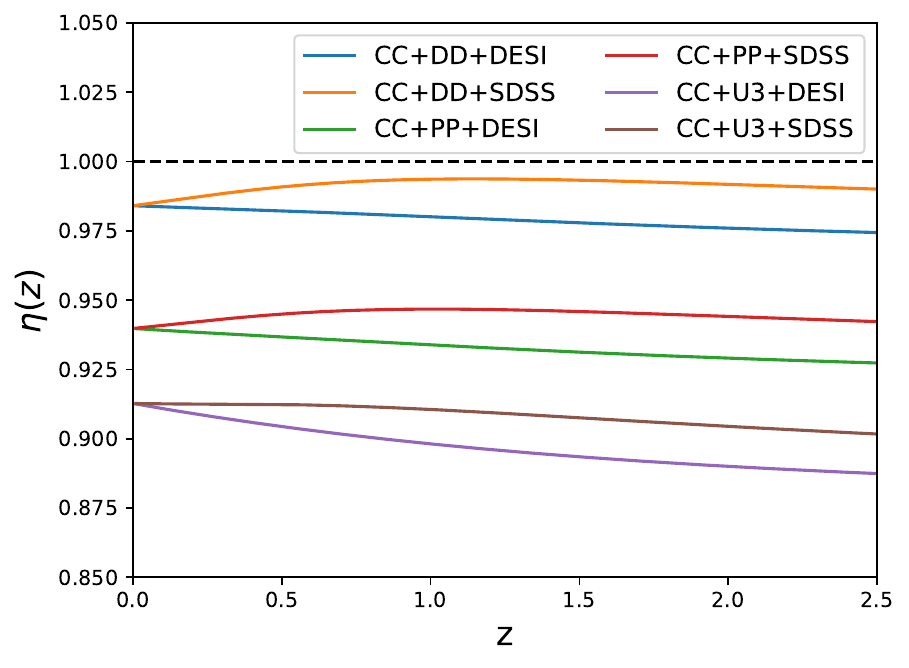}
    \caption{Theoretical $\eta(z)$ from best-fit $\Lambda$CDM parameters of individual CC, BAO, and SNe~Ia datasets.}
    \label{fig:eta_theory}
\end{figure}

We further construct theoretical $\eta(z)$ curves (Fig.~\ref{fig:eta_theory}) using the best-fit $\Lambda$CDM parameters from separate MCMC fits to the CC \(H(z)\), BAO \(R_{MH}(z)\), and SNe~Ia \(D_L(z)\) data. 
Here \(r_d\) has been eliminated through the ratio \(R_{MH}\), and the SNe~Ia fits use the published \(\mu\) directly, without additional \(M_B\) assumptions.
These curves deviate from $\eta=1$ precisely because the three datasets prefer different $H_0$ values, and the resulting spread closely resembles the differences among the SNe~Ia compilations seen in the GP and ANN reconstructions. 
The BAO curves differ only through their preferred $\Omega_m$ values, since $R_{MH}$ is independent of both $r_d$ and $H_0$, but this difference is modest. 
We therefore focus on the $H_0$-driven differences among the SNe~Ia compilations.

The correspondence between $\delta H_0/H_0^{\rm CC}$ and $\delta\eta$ indicates that the $\eta\neq1$ signal is driven by the $H_0$ tension rather than by new physics violating the CDDR. 
Because our CDDR test is designed to be independent of $r_d$, the only remaining parameter that enters as an external input is the SNe~Ia absolute magnitude $M_B$ via $\mu=m_b-M_B$. 
Since $H_0$ cannot be directly adjusted and is degenerate with $M_B$~\cite{Dhawan:2022gac, Freedman:2021ahq, Li:2025htp, Matthewson:2024ffb}, varying $M_B$ provides a direct probe of how the $H_0$ tension propagates into $\eta(z)$. 

\begin{figure}[htbp]
    \centering
    \includegraphics[width=0.75\linewidth]{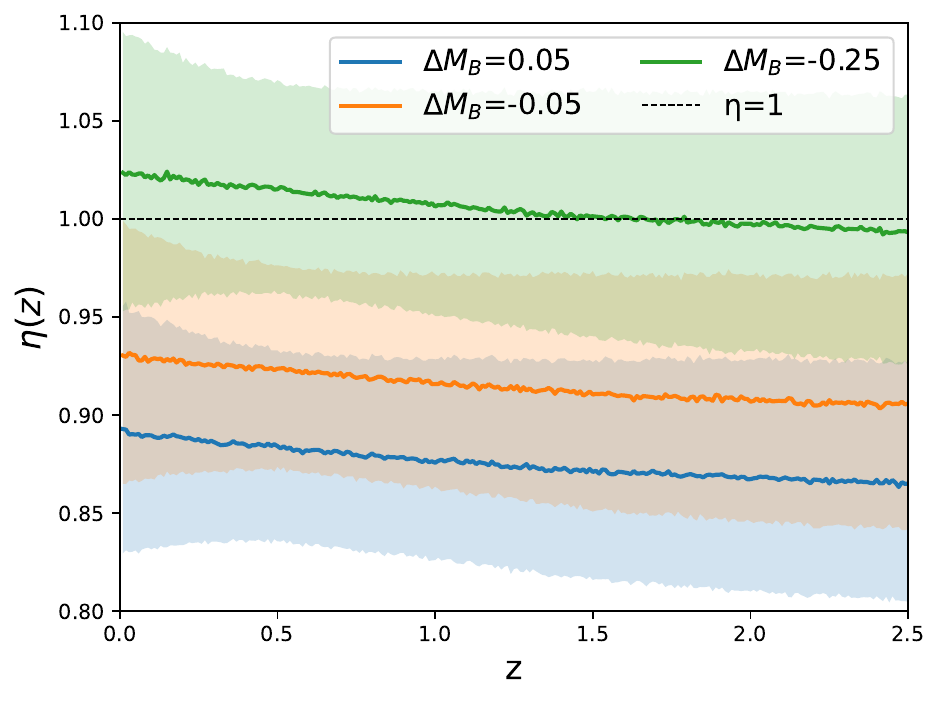}
    \caption{Reconstructed $\eta(z)$ (CC+DESI+U3) for different $\Delta M_B$ values. Decreasing $M_B$ shifts $\eta(z)$ upward.}
    \label{fig:eta_mb}
\end{figure}

To demonstrate this, we vary $M_B$ for the CC+DESI+U3 combination, which exhibits the largest deviation. 
Figure~\ref{fig:eta_mb} shows the reconstructed $\eta(z)$ for $\Delta M_B = +0.05$, $-0.05$, and $-0.25$ mag. 
A clear systematic trend emerges: decreasing $M_B$ shifts $\eta(z)$ upward. 
Table~\ref{tab:H0_MB} lists the corresponding $H_{0,\rm hyper}$ posterior constraints. 
As $M_B$ decreases, the inferred $H_0$ drops from $77.3$ to $67.4$ km/s/Mpc, approaching the CC-only value of $68.9$ km/s/Mpc. 
Varying $M_B$ is equivalent to shifting the effective $H_0$; a smaller $M_B$ reduces $\delta H_0$, and $\eta$ moves toward unity. 
This demonstrates that the $\eta\neq1$ signal is transmitted through the $M_B$--$H_0$ degeneracy, and is therefore more likely a consequence of $M_B$ calibration systematics than a true violation of the CDDR.

\begin{table}
    \centering
    \begin{tabular}{ccc}
    \hline
    Dataset & \(\Delta M_B\) & \(H_{0,\rm hyper}\) / [km/s/Mpc] \\[1ex]
    \hline
    \\[-1.8ex]
    CC-only & --- & \(68.9^{+4.5}_{-4.2}\) \\[1ex]
    \hline
    \\[-1.8ex]
    \multirow{4}{*}{Union3} & \(0\) & \(75.7^{+3.2}_{-3.1}\) \\[1ex]
     & \(+0.05\) & \(77.3^{+3.4}_{-3.1}\) \\[1ex]
     & \(-0.05\) & \(74.1^{+3.2}_{-3.3}\) \\[1ex]
     & \(-0.25\) & \(67.4^{+2.8}_{-2.6}\) \\[1ex]
    \hline
    \end{tabular}
    \caption{Posterior constraints on the mean-function hyperparameter $H_{0,\rm hyper}$ from Gen GP reconstructions of CC-only and Union3 under different $M_B$ assumptions.}
    \label{tab:H0_MB}

\end{table}

Recent studies have increasingly focused on the role of $M_B$ in cosmological analyses~\cite{Dhawan:2022gac, Freedman:2021ahq}. 
Hu et al.~\cite{Hu:2026yda} simultaneously constrained the CDDR violation parameter and a possible redshift evolution of $M_B$, parameterized as $M_B(z)=M_0+\epsilon z$, finding no statistically significant evidence for such evolution. 
Teixeira et al.~\cite{Teixeira:2025czm} explored whether CDDR violation could explain the Hubble tension and assessed its impact on evidence for dynamical dark energy. 
Matthewson and Shafieloo~\cite{Matthewson:2024ffb} examined the consistency of $M_B$ calibrations across different SNe~Ia compilations. 
Other studies have adopted fixed-$M_B$ strategies similar to ours~\cite{Li:2025htp, Kanodia:2025jqh, Zhang:2025qbs}, or explored the connection between $M_B$, $H_0$ tension, and dark energy dynamics~\cite{vonMarttens:2025dvv, Wang:2026kor}. 
The clear $M_B$ dependence highlights the need for independent SNe~Ia calibration methods, such as the tip of the red giant branch (TRGB)~\cite{Dhawan:2022gac, Freedman:2021ahq}, and a deeper investigation of unaccounted local systematics~\cite{Matthewson:2024ffb, Hu:2026yda, Tiwari:2026pzk}.

\section{Conclusion}
\label{sec:conclusion}

In this work, we have presented a model-independent test of the CDDR that is also independent of \(r_d\) calibration. 
By combining CC \(H(z)\) with the BAO distance ratio \(R_{MH}=D_M/D_H\), we obtain \(D_A(z)\) independently of \(r_d\). 
To reconstruct $H(z)$, $R_{MH}(z)$, and $D_L(z)$, we employ two independent non-parametric methods: Gen GP, with full Bayesian marginalization over hyperparameters to avoid dependence on the mean function and kernel, and ANN as a purely data-driven cross-check.

Our results show no evidence for a violation of the CDDR exceeding \(2\sigma\) over the redshift range \(0<z<2.5\). 
For the CC+BAO+DES-Dovekie combination, \(\eta=1\) lies within the \(1\sigma\) confidence interval over most of this range, with only a slight excess in certain intervals. 
The mild \(\eta\neq1\) signal observed in the other combinations is closely tied to the choice of SNe~Ia compilation: U3 yields the largest deviation (\(\sim 10\%\)) and DD the smallest (\(\sim 2\%\)). This feature matches the \(H_0\) tension among these datasets, and varying the absolute magnitude \(M_B\)---the only external parameter remaining after \(r_d\) is eliminated---systematically shifts $\eta(z)$ toward or away from unity. The effect is transmitted through the $M_B$--$H_0$ degeneracy, indicating that the $\eta\neq1$ signal is more likely a consequence of $M_B$ calibration systematics than a true violation of the CDDR.

The comparison between Gen GP and ANN reveals the impact of data sparsity on purely data-driven reconstructions. ANN exhibits a broad peak and subsequent downturn in \(\eta(z)\) that are traced to its reconstruction of CC \(H(z)\); replacing this with the Gen GP reconstruction suppresses these behaviors. The low-redshift behavior, by contrast, depends on the BAO sample, with the addition of even a single extra BAO point tightening the constraints. These tests point to future high-redshift CC measurements and low-redshift BAO data as key ingredients for more precise CDDR tests.

Further model-independent and non-parametric tests of the CDDR will benefit significantly from future high-quality data with increased statistics and wider redshift coverage. 
DESI has recently completed its five-year survey, and its upcoming data releases will provide more BAO measurements, particularly at the low redshifts. 
On the SNe~Ia side, improved \(M_B\) calibration from independent distance-ladder measurements such as TRGB or multi-indicator networks will directly test whether the \(\eta\neq1\) signal is driven by \(H_0\) tension or \(M_B\) systematics. 
More broadly, methods that directly provide \(\mu\) or \(D_L\) without \(M_B\) calibration extend the calibration-free philosophy of our \(r_d\)-free approach to the luminosity distance.

\acknowledgments
This work is supported by the National Natural Science Foundation of China Grant No. 12575063. Part of the numerical calculations in this paper have been done on the supercomputing system in the Supercomputing Center of Wuhan University. 

\bibliography{main}

\end{document}